%% file: main.tex
\documentclass{article}  
\usepackage{fullpage}
\usepackage{graphicx} % more modern
\usepackage{subfigure}
\usepackage{hhline}

\usepackage{algorithm}
\usepackage{algorithmic}
\usepackage{enumerate}
\usepackage{multirow}
\usepackage{wrapfig}
\usepackage{graphicx}
\usepackage{subfigure}
\usepackage{url}
\usepackage{makecell}
\usepackage{amsmath}
\usepackage{amssymb}
\usepackage{appendix}

\usepackage{color}

\newcommand{\zv}{\mathbf{z}}

\newcommand{\Zv}{\mathbf{Z}}

\newcommand{\wv}{\mathbf{w}}
\newcommand{\Wv}{\mathbf{W}}

\newcommand{\alphav}{\boldsymbol \alpha}
\newcommand{\betav}{\boldsymbol \beta}
\newcommand{\thetav}{\boldsymbol \theta}
\newcommand{\Thetav}{\boldsymbol \Theta}

\newcommand{\phiv}{\boldsymbol \phi }
\newcommand{\Phiv}{\boldsymbol \Phi }

\usepackage{hyperref}

\newcommand{\ericw}[1]{{\color{blue}{#1}}}

\newcommand{\st}[1]{} % remove strike-through's
\renewcommand{\ericw}[1]{{\color{black}{#1}}}

\title{Model-Parallel Inference for \ericw{Big} Topic Models}

\author{
Xun Zheng\thanks{School of Computer Science, Carnegie Mellon University, USA 15213. Email: \{xunzheng,jinkyuk,epxing\}@cs.cmu.edu},
Jin Kyu Kim$^\ast$, 
Qirong Ho\thanks{Institute for Infocomm Research,
A*STAR,
Singapore 138632.
Email: hoqirong@gmail.com}, 
Eric P. Xing$^\ast$
}
\date{}

\begin{document}

\maketitle

\begin{abstract}
In real world \ericw{industrial} applications of topic modeling, the ability to capture gigantic conceptual space \ericw{by learning an ultra-high dimensional topical representation, i.e., the so-called ``big model'',} is becoming the next desideratum after enthusiasms on "big data", 
especially for fine-grained downstream tasks such as online advertising, \ericw{where good performances are usually achieved by regression-based predictors built on millions if not billions of input features.} 
%\ericw{Conventional big-data-driven topic models only extracts thousands of topics from tens of billions of documents, which are, albeit appropriate for human interpretation, fall short of meeting the needs from such industrial ultra-high dimensional predictive machines.} 
\st{However} \ericw{The conventional data-parallel approach for training gigantic topic models} turns out to be rather inefficient in utilizing the power of parallelism, due to the heavy dependency on a \ericw{centralized image of} ``model''.
% -- be it model variables, sufficient statistics, or any shared states between processes; synchronization speed \st{of the large model} can be greatly affected by the network conditions, \ericw{especially in case of a "big model".}
Big model size also poses another challenge on the storage, where available model size is bounded by the smallest RAM \ericw{of nodes}.
To address these issues, we explore another type of parallelism, namely {\it model-parallelism}, which \ericw{enables training of} disjoint blocks of \ericw{a big topic} model in parallel. 
By integrating data-parallelism with model-parallelism,
%taking advantage of both model-parallelism and data-parallelism, 
we show that dependencies \ericw{between distributed elements} can be handled seamlessly, achieving not only faster convergence but also an ability to tackle \ericw{significantly bigger} model size. We describe an architecture for model-parallel inference of LDA, and present a variant of collapsed Gibbs sampling algorithm tailored for it. 
%Experimental results demonstrate the promise.
Experimental results demonstrate the ability of this system to handle topic modeling with unprecedented amount of 200 billion model variables only on a low-end cluster with very limited computational resources and bandwidth. 
\end{abstract}                                                                             

\input{intro}

\input{lda}

\input{parallel}

\input{gibbs}

\input{exp}

\input{conclusion}

\end{document}

%% file: intro.tex
%!TEX root = main.tex

\section{Introduction}
Recent advances in storage and network technology have brought a new era of ``big data'', where efficient processing and \st{learning} \ericw{distillation} of massive datasets has become one of the major pursuits in the field of machine learning (ML). Numerous algorithms and systems have been proposed to scale up ML for various tasks. 
However, many of the \st{proposed artifacts} \ericw{existing systems positioned for Big ML} such as MapReduce~\cite{Dean08} or Spark~\cite{Zaharia10} resort to data-parallelism, based on the assumption that the tasks associated with \ericw{partitions of the} data are independent, \ericw{and/or pose only mild reliance on synchronization.} Such assumptions are indeed \ericw{valid} for a majority of the ``data processing'' tasks such as keyword extraction from huge log files \ericw{or conventional database operations that typically sweep the data only once.}
\st{
Yet, }
Different from traditional data processing, many machine learning algorithms are not well suited for data-parallelism, due to the coupling of distributed elements through ``shared states'' \ericw{such as model parameters, latent variables, or other intermediate states; for simplicity, we refer to such entities as the ``model'' underlying the data.} \ericw{A clear dichotomy between {\it data} (which is conditionally independent and persistent throughout the process of training) and {\it model} (which is internally coupled, and is transient before converging to an optimum), and the needs for an iterative-convenient procedure to learning the model from the data is the hallmark of machine learning programs.}  
For example, \ericw{in the LDA topic model}~\cite{Blei03}, \ericw{the model to be extracted from the data} \st{as a latent subspace model} consists of a large collection of subspace bases, i.e., latent topic vectors, which are shared among all the documents; \ericw{for each of such bases, elements thereof are coupled by normality and non-negativity constraints; and estimators of all such bases bear no close-form and must be approximated through some iterative procedures. All these render any trivial parallel treatment of model and data elements impossible.}

To \st{circumvent the dependency that accompanies with} \ericw{achieve efficient distributed topic modeling under dependency constraints via a data-parallel scheme}, the following approaches have been commonly considered.
1) \ericw{Exploring approximate independencies among sub-tasks:} \st{Seek for independent tasks.}
For example in \cite{Nallapati07, Zhai12} the variational inference algorithm is decomposed into independent subtasks and parallelized.
This \ericw{strategy} in fact amounts to a Bulk Synchronous Parallel (BSP) computation model, \st{where} \ericw{of which} the drawback is evident: too many synchronization barriers \ericw{needed to ensure logical correctness} can result in \st{more} \ericw{a large number of} idle cycles and therefore is hard to achieve high scalability. 
2) Fine grained locking: 
 One can employ various locking mechanisms on shared variables to prevent the reader-writer's problem. However is it only viable on the shared memory settings. 
An early version of GraphLab~\cite{Low12} engine implemented a sophisticated version of fine grained locking mechanism in distributed settings, however at the expense of encoding the model into a graph.
3) Brute-force parallelization: In this case no specific action is taken to prevent error from  being generated during the asynchronous updates.
Some early attempts \cite{Newman07, Asuncion08}, current state-of-the-art distributed LDA inference method Yahoo!LDA~\cite{Ahmed12}, and recent advances in parameter server~\cite{Ho13,Li13} can be viewed as instantiations of the mechanism to some extent.
The major problem with this approach is that there is \st{few} \ericw{little} guarantee on the correctness of the inference procedure.
Although recent studies\cite{Recht11, Johnson13} have shown some justifications for this approach, for now the theory only supports simple models 
%\ericx{give some examples here} 
(e.g., Gaussians~\cite{Johnson13})
or requires certain assumptions 
%\ericx{mention some assumptions here} 
(e.g., updates are not overlapping too much~\cite{Recht11})
to hold as well.
Empirically, as we show later, the convergence speed of such error-prone parallelization can be improved significantly \ericw{if one can}  eliminate the parallelization error.

Apart from the dependency issue, large-scale topic modeling also \ericw{poses a challenge on how to accommodate and handle} \st{features the problem of} gigantic model size, which \ericw{has} received less attention in the literature.
Unlike \ericw{in} academic convention,
industry-scale applications of topic modeling, for instance online advertising, typically go beyond extracting only topics for human interpretation or visualization, \ericw{and feature a need for ultra-high scales on vocabulary size and topic dimensions.}
However,
most data-parallel schemes implicitly assume an image of all shared model states are readily available in each worker process, 
since it can be extremely expensive to fetch \st{variables} \ericw{them} from remote processes during \ericw{iterative training steps.} \st{the inference.}
Such assumption of having a local copy of the model breaks down when facing the big model problems.
For example modeling a corpus with a vocabulary of $10^7$ terms in a $10^5$ dimensional latent space would require $10^{12}$ model variables to be estimated. 
In real-world applications this is not uncommon considering the feature augmentation (e.g., taking word combinations) and large conceptual space behind the text.
Since the raw model may already take terabytes of storage, 
unstructured data-parallel approach is unlikely to be applicable.

To address these issues, we take advantage of a different type of parallelization mechanism, namely {\em model-parallelism}, to complement data-parallelism.
Originated from a machine learning perspective, model-parallelism addresses the above problems by carefully scheduling the updates based on dependencies \ericw{among model states} induced by the inference algorithm.
Specifically, we make use of the fact that in Gibbs sampling for LDA, the shared state access is limited to a small subset of the entire model during the \ericw{computation of an} update \st{for} \ericw{from} a data sample.
In other words, if the subsets are small enough, it is possible to find a class of disjoint subsets whose updates are completely independent of each other.
Based on the iid assumption, parallelizing over the disjoint blocks produces exactly the same result as the serial execution.
Thus model-parallel inference not only ensures the inference quality, but also reduces memory requirement by partitioning both the data and the model space.
In fact, we demonstrate the ability to handle topic modeling with 200 billion model variables on a cluster of 64 low-end machines.
We also note that the model-parallelism is suitable not only for LDA but also for many more machine learning programs. 
Primitives for more general model-parallelism can be found in \cite{Lee14}.
 
{\bf Related works}: 
Various methods have been proposed to enable large scale inference for topic models. 
In \cite{Zhai12}, a MapReduce based parallelization is presented, by making use of the independent tasks in variational inference algorithm for LDA. 
The current state-of-the-art distributed inference for LDA \cite{Ahmed12} resorts to fast background synchronization of the model. 
A rotation-scheduling idea has been studied in \cite{Yan09}, however we target distributed settings where things become more challenging due to network latency and smaller degree of parallelism (compared to GPU).
GraphLab~\cite{Low12, Gonzalez12} LDA application can be seen as a special case of model-parallelism, where by definition of the graph only non-overlapping subgraphs (i.e., documents and words) are processed simultaneously. 
Recent study on streaming variational Bayes~\cite{Broderick13} also proposed a distributed inference algorithm, however specialized in the single-pass scenario. 

%{\bf Outline}: 
\ericw{Here is an outline of the rest of the paper:}
we begin with a briefly introduction of the LDA model and the collapsed Gibbs sampling algorithm in section~\ref{sec:lda}. 
Then in section~\ref{sec:mp} we present the big picture and motivation of model-parallel inference for LDA, whose technical details are shown in section~\ref{sec:fs}. 
Distributed experiments are conducted in section~\ref{sec:exp}, and finally section~\ref{sec:end} concludes.
%\ericx{For experiments, I wonder if we want to show }

%% file: lda.tex
%!TEX root = main.tex

\section{Latent Dirichlet Allocation}\label{sec:lda}
Latent Dirichlet allocation (LDA)\cite{Blei03} is a hierarchical Bayesian topic model that learns a low-dimensional representation for a high-dimensional corpus.
Because of the ability of capturing latent semantics underlying the text,
it is widely applied to various real world tasks such as online advertising and personal recommendation.
In recent years, with increasing amount of data the need for larger conceptual space is also emphasized, posing a challenge in large scale inference for LDA.
In this section, we briefly overview the LDA model definition and its inference using collapsed Gibbs sampling. 

\subsection{The Model}
LDA considers each document as an admixture of $K$ topics, where each topic $\phiv_k$ is a multinomial distribution over a vocabulary of $V$ words. 
For each document $d$ a topic proportion vector $\thetav_d$ is drawn from $\mathrm{Dirichlet}(\alphav)$.
Then for each token $n$ in the document, a word $w_{dn}$ is drawn from $\mathrm{Multinomial}(\phiv_{z_{dn}})$
and a topic assignment $z_{dn}$ is drawn from $\mathrm{Multinomial}(\thetav_d)$.
In fully Bayesian LDA, topics are random samples drawn from Dirichlet prior, $\phiv_k \sim \mathrm{Dirichlet}(\betav)$.

Given the corpus $\Wv = \{\wv_d\}_{d=1}^D$ where $\wv_d = \{w_{dn}\}_{n=1}^{N_d}$,
LDA infers  the posterior $p( \Phiv, \Thetav, \Zv | \Wv )$.
%represent the tokens appear in document $d$;
%$\Phiv = \{ \phiv_k \}_{k=1}^K$  the set of topics; 
%$\Thetav = \{ \thetav_d \}_{d=1}^D$ the topic mixture proportion of the documents;
%$\Zv = \{ \zv_d \}_{d=1}^D$ the topic assignments of all tokens in the corpus. 
%Given the corpus $\Wv$, LDA infers the posterior 
%\begin{align}\label{eqn:post}
%p( \Phiv, \Thetav, \Zv | \Wv ) = \frac{p_0(\Phiv, \Thetav) p(\Zv | %\Thetav) p(\Wv | \Phiv, \Zv)}{p(\Wv)},
%\end{align}
% where $p_0(\Phiv, \Thetav)$ denotes priors on the latent variables, $p(\Zv | \Thetav)$ and $p(\Wv | \Phiv, \Zv)$ follows from the generative process.
However exact inference is intractable due to the normalization term, hence approximation methods such as Gibbs sampling comes into play.
The mixing rate can be further accelerated by integrating out intermediate Dirichlet variables $(\Phiv, \Thetav)$ analytically, yielding the collapsed Gibbs sampling algorithm~\cite{Griffiths04}:
%\vspace{-2ex}
\begin{align}\label{eqn:cond}
p (z_{dn} = k  | \Zv_{\neg dn})
\propto 
\frac{ (C_{d, \neg n}^k + \alpha_k) (C_{k, \neg n}^t + \beta_t) }{ C_{k, \neg n} + \sum_t \beta_t}, 
\end{align}
where $t$ is the word that the current token $w_{dn}$ maps to; $C_d^k$ is the number of tokens assigned to topic $k$ in document $d$; $C_k^t$ is the number of times term $t$ has been assigned to topic $k$; $C_k$ is the total number of tokens assigned to topic $k$; and finally $C_{\cdot, \neg n}^\cdot$ is the count excluding the $n$-th token. 
	
\subsection{Sparse Sampling}
$O(K)$ time complexity for each topic assignment in collapsed Gibbs sampling still leaves room for further improvement.
In \cite{Yao09}, a sub-linear complexity sampling algorithm that makes use of sparsity is introduced.
The motivation comes from the observation that the counts $\{C_d^k\}$ and $\{C_k^t\}$ are sparse: only a few out of $K$ entries are filled with non-zero counts in every document or term. 
As we shall discuss later, similar idea is also applicable to the model-parallel inference.

The fast sampling algorithm starts with decomposing the conditional~\eqref{eqn:cond} in the following way:
\begin{align}\label{eqn:fast}
p (z_{dn} = k | \Zv_{\neg dn}) 
\propto \mathbf{A}_k + \mathbf{B}_k + \mathbf{C}_k,
\end{align}
where
\begin{align*}
\mathbf{A}_k = \frac{\alpha_k \beta_t}{C_{k, \neg n} + \sum_t \beta_t} ,
\quad
\mathbf{B}_k = \frac{\beta_t}{C_{k, \neg n} + \sum_t \beta_t}\ C_{d, \neg n}^k ,
\quad
\mathbf{C}_k = \frac{\alpha_k + C_{d, \neg n}^k }{C_{k, \neg n} + \sum_t \beta_t}\ C_{k, \neg n}^t .
\end{align*}

Since $\mathbf{A}_k$ is dense, $\sum_k \mathbf{A}_k$ can be precomputed in $O(K)$ and maintained in $O(1)$ time; 
$\sum_k \mathbf{B}_k$ can be precomputed in $O(K_d)$ time where $K_d$ is the average number of nonzero entries in $\{C_d^k\}$;
the fractional term in $\mathbf{C}_k$ can be precomputed in $O(K)$ time
and $\sum_k \mathbf{C}_k$ can be constructed in $O(K_t)$ time by taking advantage of sparsity in $\{C_k^t\}$, whose expected sparsity is $K_t$. 
Note that not only the construction of conditional distribution takes sub-linear complexity in $K$, the sampling procedure can also benefit significantly from the sparsity of $\{C_k^t\}$ and $\{C_d^k\}$, due to the observation that $\mathbf{B}$ and $\mathbf{C}$ bucket contain most of the probability mass. 
So the overall time complexity for sampling one topic assignment $z_{dn}$ is $O(K_d + K_t)$.

%% file: parallel.tex
%!TEX root = main.tex

\section{Model-Parallel Inference}\label{sec:mp}
Data-parallel inference for LDA typically distributes different set of documents to the workers to perform Gibbs sampling, while sharing a central model across all of them via some synchronization schemes.
Albeit being powerful in handling large amount of data, it introduces two potential issues that are less recognized in the literature.
1) It fails to handle huge model. 
In data-parallel inference, it is natural to assume that the entire copy of the model is available in the workers throughout the inference procedure.
However as we mention earlier the need for big model breaks down this assumption: a model with billions of variables can easily exceed the reasonable RAM size these days.
2) It cannot control inconsistency in the shared model.
Most data-parallel inference trades correctness with performance.
For example in \cite{Ahmed12}, the shared model is updated by a separate thread cycling over the local model, hoping for the inconsistency does not affect the algorithm by much. 
However this strategy relies heavily on the network condition, as we show in Section~\ref{sec:exp}: 
for low bandwidth networks, the effect of inconsistency becomes evident since the algorithm proceeds without noticing the slow synchronization in the background.
%\eric{It would be nice to provide experimental evidence of this claim. Maybe no time for tomorrow's deadline, but better do it right after submission to make it in time for the rebuttal.}

%Noticing these drawbacks of data-parallel inference, we demonstrate our solution to the problems in this section.

\subsection{Dynamic Model Partitioning}
A monolithic treatment of shared model in data-parallel inference often fails to address the ``big model'' problem,
which can take place when 1) massive number of the model variables or parameters are introduced by the statistical model; or
2) huge additional shared data structure is required to assist the inference algorithm.  
In either case, having a complete copy of the model in every worker is a potential danger not only because it may fail to load the model in the first place, but also because adding computing nodes will not help reduce the memory consumption of individual workers.

Our solution to this issue is to partition the shared model into disjoint blocks.
This is motivated by the fact that each step of Gibbs sampling only requires change in a small subset of the entire statistics $\{C_k^t\}$,
hence certain degree of parallelization can be achieved on the model side, in addition to the data.
Specifically, since Gibbs sampling for two distinct words are nearly independent\footnote{We discuss the dependency on $ \{C_k\} $ later.}, the $V \times K$ word-topic count matrix can be effectively partitioned by words.
The outcome of model partitioning is straightforward: it reduces the model size on each workers, and also achieves scalability on the model by allowing more nodes to share the burden. 
Note that dynamic model partitioning is a complement to data partitioning rather than a replacement.
Instead of static placement, it provides more flexibility to the algorithm and ensures each worker to work on the complete set of model during the inference, rather than only a subset of them.

\begin{algorithm}[t]
\caption{Scheduler}\label{alg:sched}
\begin{algorithmic}[1]
  \STATE {\bf Initialization:} construct the task pool.
  \WHILE{true}
  	\STATE dispatch tasks to workers;
  	\STATE rotate tasks;
  \ENDWHILE
\end{algorithmic}
\end{algorithm}

In model-parallel LDA, dynamic partitioning of the model is realized by a {\em scheduler} component, as described in Algorithm~\ref{alg:sched}.
Specifically,  it first divides the $V$ words into $M$ disjoint blocks $\{ V_1, V_2, \dotsc, V_M \}$.
Each block $V_m$ is assigned to corresponding worker $m$ as the initial set of {\em tasks}.
Therefore each worker $m$ only samples tokens $z_{dn}$ such that $w_{dn} \in V_m$. 
Once all the workers have finished sampling their own blocks, the scheduler rotates the blocks to different workers for another round (sub-iteration) of sampling:
worker $m$ acquires the block $V_{m'}$ where $m' = (m + 1) \text{ mod } m$.
After $M$ rounds of sampling, all topic assignments $\Zv$ will have been sampled exactly once.
This amounts to an {\em iteration} over the data and we repeat the process until convergence. 

\subsection{On-demand Communication}
Synchronization of shared model is another major issue in data-parallel inference.
Existing methods such as \cite{Ahmed12} has been mainly focused on efficient maintenance of the word-topic count matrix $\{C_k^t\}$, for example using an asynchronous key-value store to frequently incorporate and distribute updates committed by the workers.
However, best-effort synchronization only guarantees eventual consistency, 
hence the workers may construct incorrect distributions from the staled statistics. 
The effect of staleness becomes even evident with low network bandwidth, which is common in low-end clusters and custom cloud services.
If the shared states cannot be synchronized in time even though the network is saturated, then parallelization error will only increase as the inference algorithm proceeds.

\begin{algorithm}[t]
\caption{Worker}\label{alg:worker}
\begin{algorithmic}[1]
  \WHILE{not converged}
  	\STATE receive tasks from scheduler;
  	\STATE request model blocks from kv-store;
  	\STATE Gibbs sampling using \eqref{eqn:fastxy};
  	\STATE commit new model blocks to kv-store;
  \ENDWHILE
\end{algorithmic}
\end{algorithm}

Based on dynamic model partitioning, we can avoid such issues easily by carefully managing communication between workers.
To achieve this, we introduce 
a {\em key-value store} that stores the global model $\{C_k^t\}$.
Note that different from being a ``parameter server''~\cite{Ahmed12}, the purpose of this component is mainly for distributed in-memory storage: thanks to dynamic model partitioning, frequent background asynchronous communication is no longer required.
In practice a simple distributed hash table implementation suffices the need.
Given the dynamic model partitioning strategy, on-demand communication between workers and key-value store follows the procedure described in Algorithm~\ref{alg:worker}.
At the beginning of each round, 
after receiving the task list,
each worker can start requesting its model blocks from the key-value store. 
Similarly after finishing the tasks,
workers can commit changes in local model blocks, thereby updating the global model.
This process can be further accelerated by overlapping sampling procedure and communication, i.e., send/receive model blocks asynchronously.

Again since the model blocks are non-overlapping, there is no synchronization issue on the key-value store. 
Moreover the amount of communication is reduced significantly, compared to the frequent synchronization approach.
By combining dynamic model partitioning and on-demand communication via key-value store, variable dependency between workers is also reduced. 
It not only eliminates the need for a frequently synchronized shared states as in data-parallel inference, but also results in faster sampler convergence per token processed.
In fact as we show later our method requires much fewer iterations to converge than others, while having similar per-iteration time complexity.

\subsection{Non-separable Dependency}
So far we have deliberately omitted another source of dependency, the global topic count vector $\{C_k\}$. 
It is impossible to divide $\{C_k\}$ into disjoint blocks since the term is required in sampling for all the tokens.
However, noticing the fact that the value is relatively large since $C_k = \sum_t C_k^t$ and it only appears in the denominator,
changes in small magnitude will not affect the final distribution much.
This demands for a much relaxed level of consistency.
Therefore we synchronize $\{C_k\}$ across the workers at the beginning of each round through the key-value store.
It is highly efficient since every worker only needs to send/receive a vector of size $K$ to/from the key-value store. 
During the round, workers are not aware of the changes in $\{C_k\}$ made by other workers, which causes some error in the distribution to sample from.

This is in some sense similar to the idea used in \cite{Newman07}, where the entire model is allowed to go out-of-sync during an iteration. 
However we only relax the consistency requirement on $\{C_k\}$, while the major element of the model, $\{C_k^t\}$, is maintained without any error.
As we will show in Section~\ref{sec:exp}, due to the small amount of change compared to the actual value, the resultant error is empirically negligible.

To sum up, combining dynamic model partitioning and on-demand communication not only reduces memory load of each worker, but also avoids most of the parallelization error.
A special protocol is introduced to address non-separable dependency issue on $ \{C_k\} $, without sacrificing the inference quality.
As we show later, compared to a data-parallel method \cite{Ahmed12}, model-parallel inference takes an order of magnitude less time to converge.

%% file: gibbs.tex
%!TEX root = main.tex

\section{Implementation Details}\label{sec:fs}
In this section we provide some technical details about implementing the model-parallel inference for LDA.

\begin{figure}[t]
\centering
\includegraphics[width=0.45\textwidth]{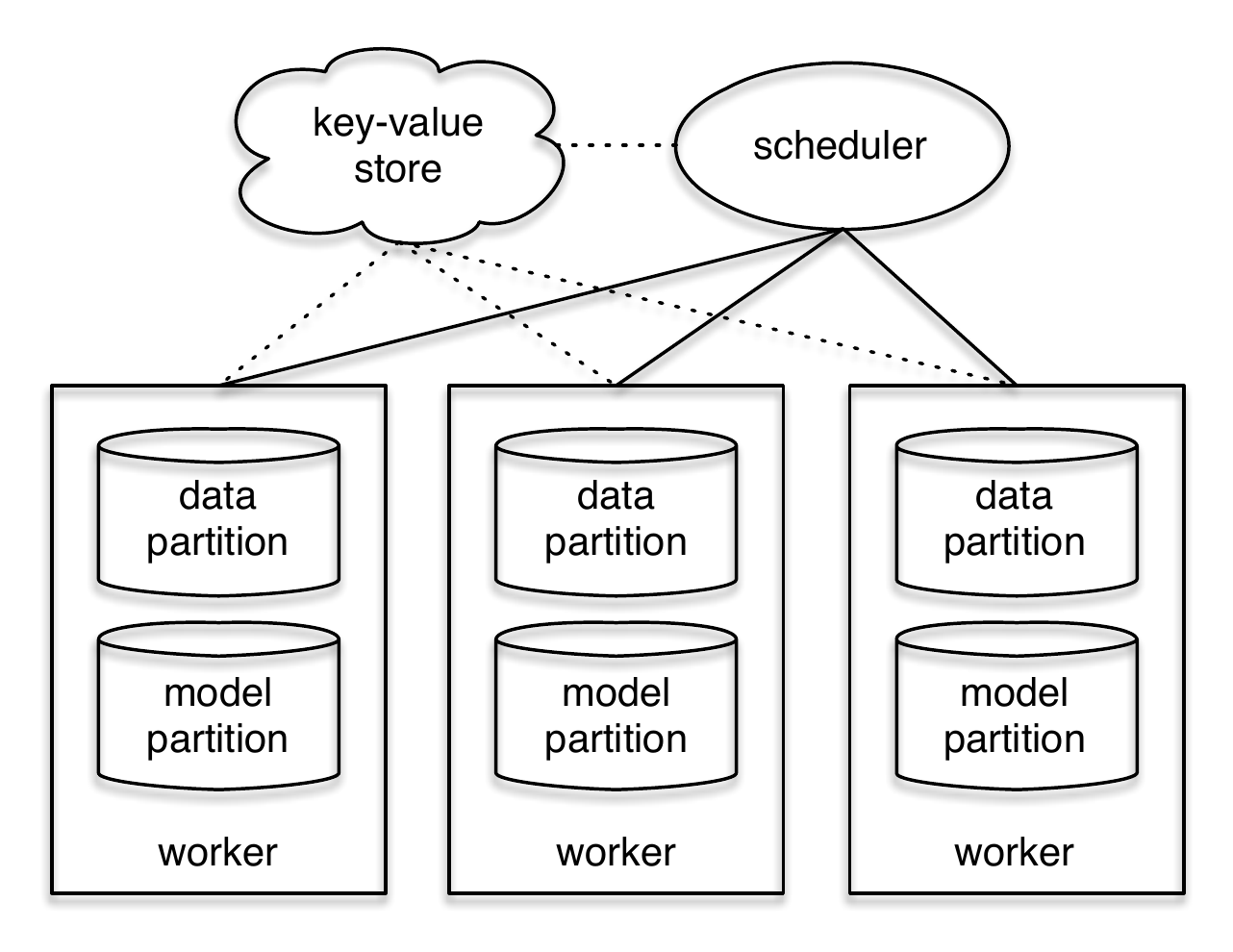}
\caption{
A high-level view of the architecture for model-parallel inference of LDA.
}  
\label{fig:arch}
\end{figure}
\subsection{Overall Architecture}
The complete design of the system components is illustrated in Figure~\ref{fig:arch}. 
We partition both data and model so that each partition can be stored in 
a single machine memory.
The scheduler directly communicates with the workers to 1) generate and assign tasks and 2) coordinate model partitions between workers. 
It also maintains a special communication channel with the key-value store to handle non-separable dependency in $ \{C_k\} $. 
As we mentioned,
model blocks are communicated via a distributed key-value store in a managed fashion, rather than busy synchronization.
This significantly reduces the amount of communication and hence lowers the requirement on  network bandwidth.

\subsection{Fast Sampling on Inverted Index}
Upon receiving the task list and necessary model blocks, 
the main job left for the worker is to perform Gibbs sampling on the local data.
Because of the scheduling constraint, only tokens that are mapped to the words in the current task list can be sampled in this round.
Traditional bag-of-words representation of the documents turns out to be rather inefficient in this case:
to determine the set of tokens to be sampled in this round, 
sequential iterations over the dataset as well as  
multiple comparisons between the task list and the token are required.

In fact this is a classic problem in search engines, where the typical solution is to represent the documents in {\em inverted index}, instead of forward index (i.e., bag-of-words).
With the inverted index created, for each worker $m$, each record indexed by word $t$ represents all the topic assignments $z_{dn}$ such that $w_{dn} = t$ and $\zv_d \in D_m$.
By doing so we can completely eliminate the multiple comparisons between two sets.

In addition, similar to the idea in \cite{Yao09}, we can take advantage of sparsity as well. 
We first note that the same decomposition~\eqref{eqn:fast} is not optimal for sampling on inverted index.
To see the reason, note that a great proportion of efficiency in the sparse sampling algorithm~\cite{Yao09} comes from precomputing $\sum_k \mathbf{B}_k$ for each document.
Once cached, additional changes to  $\sum_k \mathbf{B}_k$ within a document only requires $O(1)$ time to make incremental updates.
The caching effect is maximized when tokens in each document are sampled sequentially in a process.
However for sampling on inverted index, it is no more the case: 
$\sum_k \mathbf{B}_k$ is frequently recomputed since typically only a few tokens in a document represent a specific word.

Instead, a different decomposition can be done as follows to maximize the caching effect:
\begin{align}\label{eqn:fastxy}
p (z_{dn} = k | \Zv_{\neg dn}) 
\propto \mathbf{X}_k + \mathbf{Y}_k,
\end{align}
where
\begin{align*}
\mathbf{X}_k = \frac{C_{k, \neg n}^t + \beta_t}{C_{k, \neg n} + \sum_t \beta_t}\ \alpha_k , 
\quad
\mathbf{Y}_k = \frac{C_{k, \neg n}^t + \beta_t}{C_{k, \neg n} + \sum_t \beta_t}\ C_{d, \neg n}^k .
\end{align*}

The first probability bucket $\sum_k \mathbf{X}_k$ can be precomputed for every word (i.e., task) in $O(K)$ time, with $O(1)$ maintaining cost for future updates.
It is cached for every token associated with the word $t$ in local partition. 
Also note that the fractional terms in $\mathbf{X}_k$ and $\mathbf{Y}_k$ is identical, thus coefficients of $\{\mathbf{Y}_k\}$ can be precomputed along with $\sum_k \mathbf{X}_k$ with no additional cost.
To get $\sum_k \mathbf{Y}_k$, we  make use of sparsity in $C_d^k$ which requires $O(K_d)$ time. 
Note that due to the dense fractional term and unbiased mass partition, the algorithm is not as efficient as the sparse sampler in \cite{Yao09}.
However, as we stated above, this algorithm makes full use of the inverted index structure that is required by the model-parallel inference.
In Section~\ref{sec:exp}, we show that the disadvantage of non-optimal sampling algorithm is mitigated as the benefit of model-parallelism becomes salient.

%% file: exp.tex
%!TEX root = main.tex

\section{Experiments}\label{sec:exp}
In this section we quantitatively evaluate the proposed model-parallel inference for LDA. 
Our chosen baseline is Yahoo!LDA~\cite{Ahmed12}, which is a popular, publicly-available distributed implementation of the Sparse Gibbs sampler~\cite{Yao09}. 
Another notable baseline is Google's PLDA+~\cite{Liu11}, which has similar token sampling throughput to Yahoo!LDA --- roughly, both Yahoo!LDA and Google PLDA+ process 20K tokens per compute core, per second, on a medium-sized cluster with 10-100 machines. 
Since the sampling throughput of Yahoo!LDA and Google PLDA+ are similar, we only compare to Yahoo!LDA. 
In our experiments, we will show that our method, while having similar sampling throughput to Yahoo!LDA (and PLDA+), converges significantly faster per iteration because our careful, word-partitioned model-parallel design significantly reduces synchronization errors in the word-topic table (as in Figure~\ref{fig:error}).
We also attempted to compare with the topic modeling toolkit in GraphLab~\cite{Gonzalez12}, however in all of the experiments it failed to initialize due to excessive memory consumption. 
This is acceptable since it is an application built on top of the general-purpose system rather than a performance-driven instantiation of the algorithm, hence we omit the result hereinafter.

%We compare with the current state-of-the-art distributed inference method Yahoo!LDA \cite{Ahmed12}, which is based on data-parallelism.
%We do not include comparison with pLDA+~\cite{Liu11} because the reported throughput is similar to Yahoo!LDA. 
%We also attempted to compare with the topic modeling toolkit in GraphLab~\cite{Gonzalez12}, however in all of the experiments it failed to initialize due to excessive memory consumption. 
%This is acceptable since it is an application built on top of the general-purpose system rather than a performance-driven instantiation of the algorithm, hence we omit the result hereinafter.

{\bf Experiment Settings: }
To demonstrate the effectiveness over different hardware settings, we 
conduct experiments on two disparate settings~\cite{Gibson13}: 
a high-end cluster with 64-core machines.
a low-end cluster equipped with 2-core machines.
Specifically, the high-end cluster contains 10 machines connected via 40Gbps Ethernet network interface, each node equipped with quad-socket 16-core AMD Opteron 6272 (2.1GHz) and 128GB RAM.
The low-end cluster consists of 128 machines connected via 1Gbps Ethernet with dual-socket AMD Opteron 252 (2.6GHz) and 8GB RAM in each machine.
The model-parallel inference is fully implemented in C++11.
Note that although the high-end machines are NUMA nodes, for fare comparison we do not include any optimization for NUMA architecture.

{\bf Dataset: }
We use Pubmed\footnote{\url{https://archive.ics.uci.edu/ml/datasets/Bag+of+Words}} and the 3.9M document English Wikipedia
abstracts\footnote{\url{http://wiki.dbpedia.org/Downloads39\#extended-abstracts}} as our dataset.
We further construct an augmented corpus by extracting bigrams (2 consecutive tokens) from Wikipedia corpus. 
Pubmed contains 8.2M documents, $ V=141043 $ words and about 737.9M tokens.
The original Wikipedia dataset consists of $V=2.5$M unique words and 179M tokens,
while in bigram corpus there are $V=21.8$M unique phrases and $79$M occurrence of these phrases.
We note that bigram vocabulary of 21.8M is almost an order of magnitude larger than \cite{Ahmed12},
and clearly demonstrates our scalability to very large model sizes.
Our experiments used the number of topics from $K=1000$ up to $10000$, which results
in extremely large word-topic tables: $12.5$B elements in the unigram case, and $218$B elements in the bigram case.
The model size is about 60 times larger than the recent result~\cite{Ahmed12}.

%{\bf Evaluation: }
%The purpose of any inference algorithm, or a system that executes an inference algorithm, is to accurately learn the model parameters and variables in the shortest possible time, on the given training data set. For LDA Gibbs sampling algorithms (such as system and our experimental baselines), the key model parameter is the word-topic table, and the ideal notion of accuracy is Markov Chain Monte Carlo (MCMC) convergence to the true posterior distribution implied by the training data. However, measuring the distance of the MCMC chain to the true posterior is impractical, because finding the true posterior is an NP-complete problem; therefore we must resort to a surrogate measure to determine convergence speed and quality. For all systems tested, our chosen surrogate measure is the training log-likelihood on the most recent MCMC sample at iteration $i$, defined as

%\begin{align}
%\ell^{(i)}
%%& = \log \prod_{d=1}^D p(\wv_d, \zv_d ; \alphav, \betav) \\
%& = \log p(\Wv| \Zv, \betav ) p(\Zv| \alphav) \\
%& = \log \prod_{k=1}^K \frac{\Delta(C_k^{(i)} + \betav) }{\Delta(\betav)}
%\prod_{d=1}^D \frac{\Delta(C_d^{(i)} + \alphav)}{\Delta(\alphav)}, \nonumber
%\end{align}
%where $\Delta(\cdot)$ denotes multinomial Beta function.
{\bf Evaluation:}
We choose training log-likelihood as our surrogate measure of convergence because 1) LDA Gibbs samplers tend to converge to (one of many possible) local optima in the space of possible word-topic and doc-topic tables, and 2) the progress of the sampler to a local optimum correlates well with the rise, and then plateauing, of the training log-likelihood measured on the latest sample. Since the LDA Gibbs sampler is unlikely to leave a local optima once it has reached one, the algorithm can be safely terminated once the log-likelihood plateaus.
%One might ask why we did not employ test data perplexity as our surrogate. Perplexity is meant to compare different {\it models} (e.g. different flavors of LDA), in terms of how well they generalize to new data. However, our focus is on {\it inference}, not {\it modeling}; all systems/algorithms we tested perform inference for the standard LDA model, and therefore model generalization is not the issue under investigation. Moreover, because an inference algorithm learns model parameters and variables only from the training data, it is only appropriate to track its convergence as a function of the training data. Using test data perplexity introduces additional confounding factors, particularly how well each training data local optima generalizes to the test data --- this is a confounding factor because sampler algorithms are not designed to control which training optima they will eventually reach. The point is that training data log-likelihood controls for external factors better than test data perplexity, in the context of measuring inference speed and accuracy.
One might ask why we did not employ test data perplexity as our surrogate, as did by many practitioners. 
We caution that this metric is in fact improper for evaluating competing inference systems (on the same model), but suitable for evaluating goodness of different model designs (using the same inference system).
For instance, it can be used to evaluate different flavors of LDA or alternative models in terms of how well they capture training data characteristic and generalize to new data. 
However, our focus is on {\it inference quality and efficiency} on the \emph{same model}, not goodness of {\it competing models}; all systems/algorithms we tested perform inference for the standard LDA model, and therefore (the difference of) model generalization is not the issue under investigation. Moreover, because an inference algorithm learns model parameters and variables only from the training data, it is only appropriate to track its convergence as a function of the training data. Using test data perplexity introduces additional confounding factors, particularly how well each training data local optima generalizes to the test data --- this is a confounding factor because sampler algorithms are not designed to control which training optima they will eventually reach. The point is that training data log-likelihood controls for external factors better than test data perplexity, in the context of measuring inference speed and accuracy.

\begin{figure}[t]
\centering
\subfigure[]{
\includegraphics[width=0.22\textwidth]{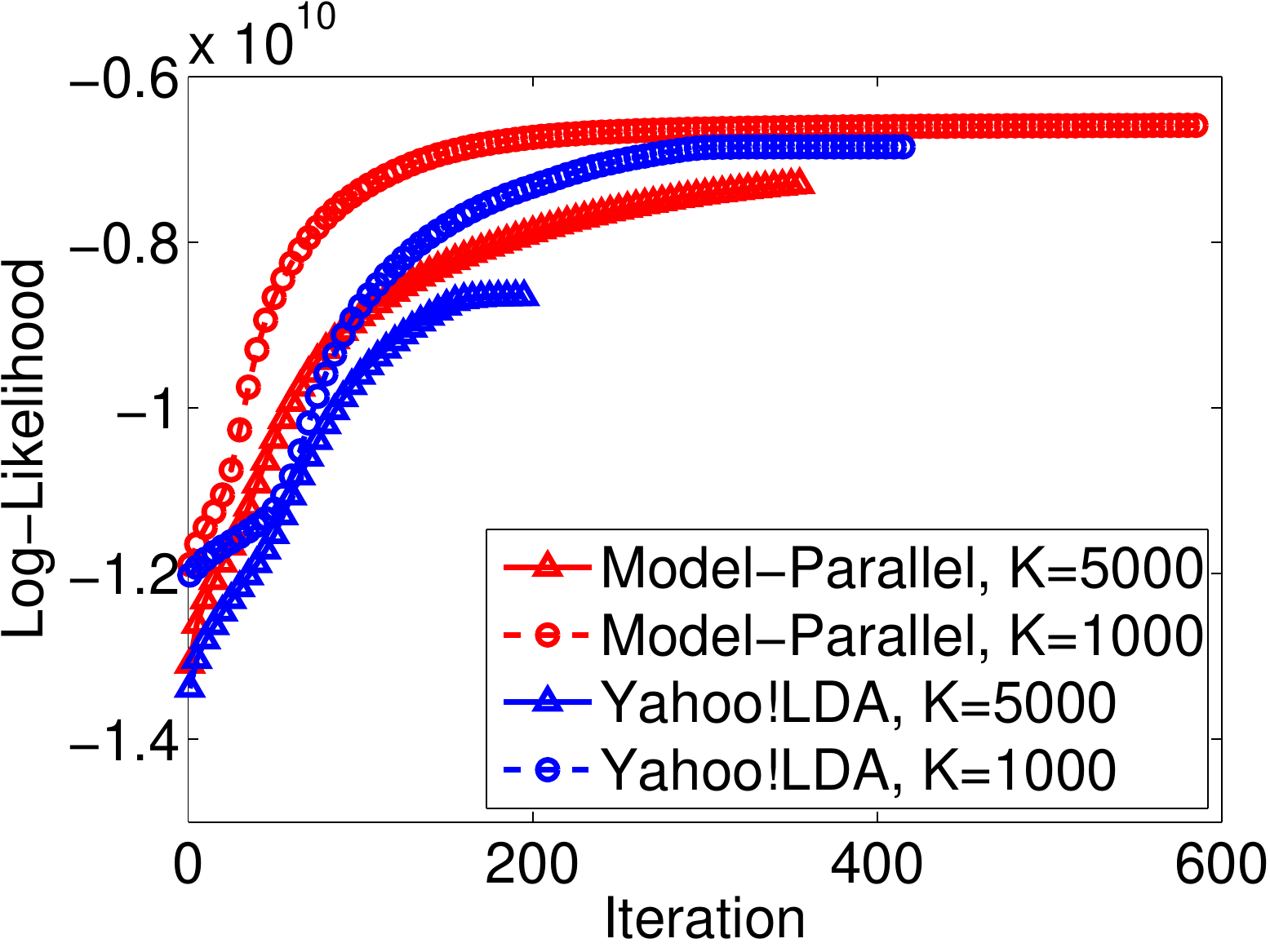}
} 
\subfigure[]{
\includegraphics[width=0.22\textwidth]{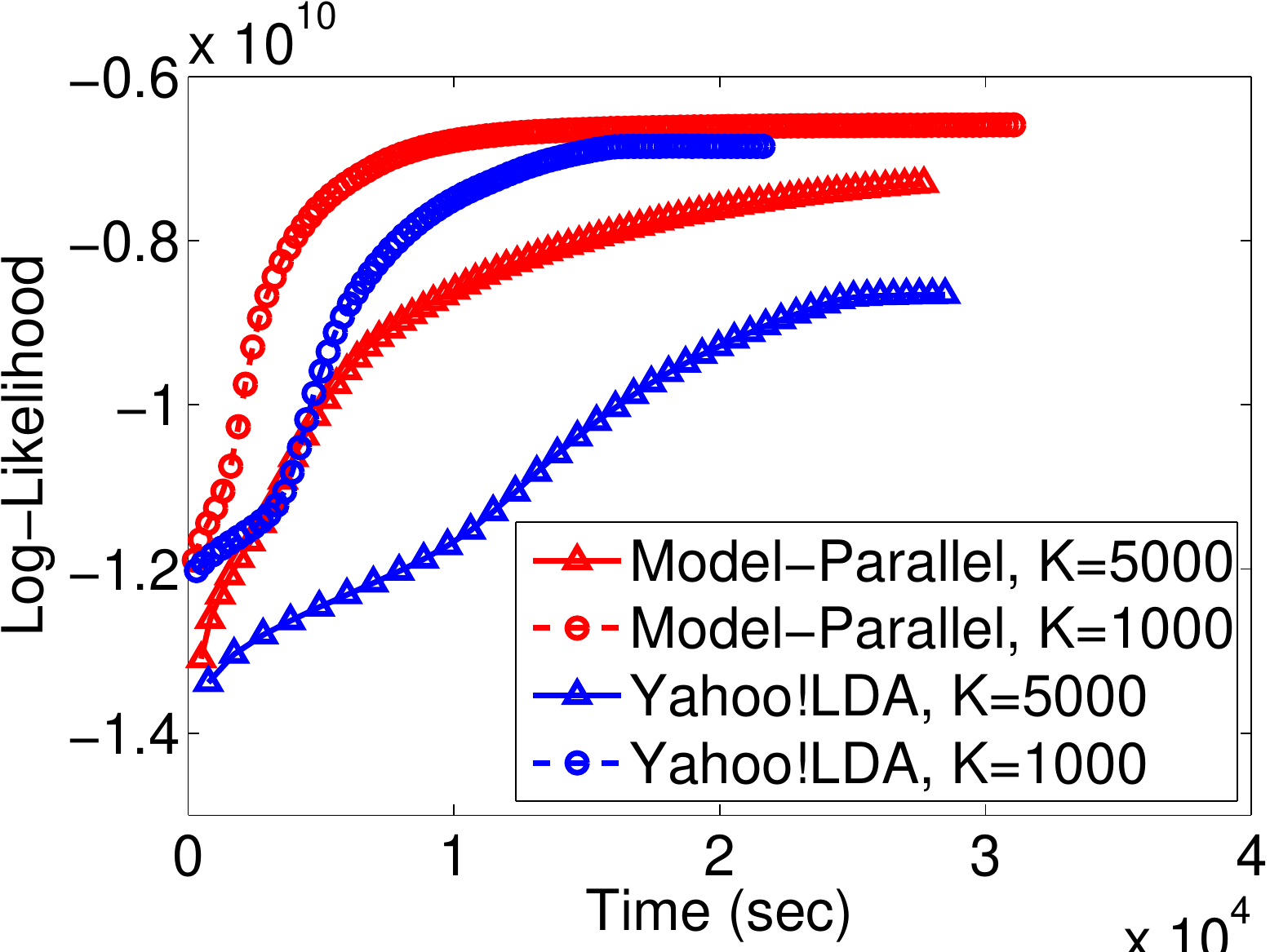}
} 
\caption{Convergence speed. Model-parallel inference exhibits sharper move toward higher likelihood.
}
\label{fig:conv}
\end{figure}

\subsection{Convergence}
We first compare convergence speed of different methods on the high-end cluster, using Pubmed dataset with 1000 and 5000 topics.
Figure~\ref{fig:conv}(a) shows the log-likelihood at each iteration. 
We can observe that model-parallel inference achieves greater per-iteration progress than data-parallel approach. 
In other words, our method requires much fewer iterations  to reach a certain likelihood.
Figure~\ref{fig:conv}(b) shows the log-likelihood trend in terms of elapsed time. 
We can observe similar trends as in per-iteration plot.
This again shows the effect of sampling from correct distributions: dynamic model partitioning seamlessly handles the dependency on the model, whereas data-parallel approach suffers slow convergence especially at the beginning due to drastic change of the model copies in worker nodes.

\begin{figure}[t]
\centering
\vspace{-4.5em}
\includegraphics[width=0.3\textwidth]{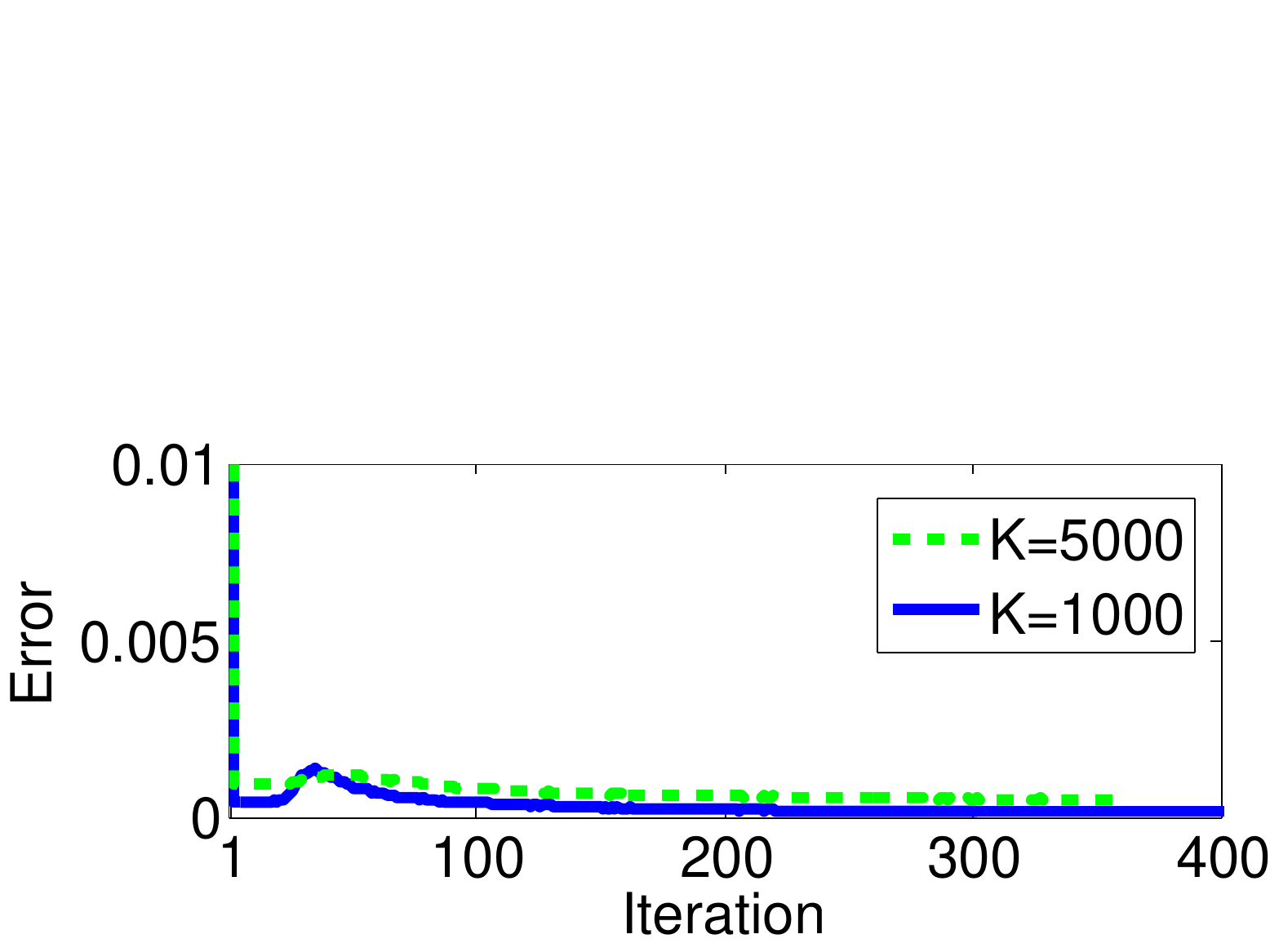}
\caption{% 
The error $\Delta_{r,i}$
at each iteration, with each round viewed as $ 1/M $ progress of an iteration.
The error is almost 0 (minimum) everywhere.
}
\label{fig:error}
\end{figure}

We also show the effect of lazy synchronization in $\{C_k\}$, which breaks down the independence between workers.
As mentioned in Section~\ref{sec:mp}, $\{C_k\}$ is only synchronized at the beginning of each round, therefore is free to go out of sync during the left of the round.
We relax the consistency requirement based on the intuition that minor change in huge counts will not affect the overall result much. 
We now show that the induced error is almost negligible in practice.

As a proxy for the error made in each round, we can measure the difference between the true $T = \{C_k\}$ and its local copy $\tilde{T}_m = \{C_k\}$ on worker $m$ at the end of each round. 
Specifically,  we define the error at each round $r$ and iteration $i$ to be
$\Delta_{r,i} = \frac{1}{M N} \sum_{m = 1}^M \| T - \tilde{T}_m \|_1,
$
where $N = \sum_k C_k$ is the total number of tokens in the corpus.
In other words, we compute the normalized $\ell_1$-distance between each worker's local copy $\tilde{T}_m$ and true value $T$, and then average the amount over all the workers. 
As a result $\Delta$ must lie in $[0, 2]$, where $0$ denotes no error.
Figure~\ref{fig:error} shows the error 
collected on high-end cluster using Pubmed dataset.
We can observe that the error 
immediately drops to 0 and stays close to it during the rest of the inference procedure.
This demonstrates that our method exhibits very small parallelization error and hence faster convergence.

\subsection{Model Size}
We demonstrate our ability to handle big models in Table~\ref{tab:bigmodel}. 
%First of all, GraphLab fails to start the algorithm in all settings. 
%This shows the overhead of the graph representation: in order to distribute a graph, a great number of vertices need to be duplicated.
Yahoo!LDA starts to fail on the problem size of 2.5M vocabulary and 10000 topics.
It is due to the fact that the local copy of the model no longer fits into the memory, even though it only stores keys that appear in the local subset of the data.
In contrast, by sharding the model into blocks, our method effectively handles bigger models.
As shown in the table, model-parallel approach is able to perform inference on all configurations of the model size, including the biggest one using bigram dataset with 10000 topics, indicating our ability to handle a model size over 200 billion only on a low-end cluster.
In addition we can observe a faster convergence in small model setting compared to Yahoo!LDA.
This indicates that model-parallelism is effective not only for big model but also for moderate-sized model problems.
All of these clearly demonstrates the effectiveness of dynamic model partitioning strategy.

\begin{table}
\centering             
\caption{Time to converge on different model size with 64 low-end machines. Model-parallel inference not only handles larger models, but also converges faster.\vspace{1ex}}\label{tab:bigmodel}
\scalebox{.95}{\begin{tabular}{|c|c|c|c|c|} 
\hline
Corpus & \multicolumn{2}{c|}{Wiki-unigram} & \multicolumn{2}{c|}{Wiki-bigram} \\ \hline
$ K $ & $5000$ & $ 10000 $ & $ 5000 $ & $ 10000 $ \\ 
\hhline{|=|=|=|=|=|}
{\bf Model-Parallel} & {\bf 2.3 hr} & {\bf 5.0 hr} & {\bf 8.9 hr} & {\bf $ > $ 12 hr*} \\ \hline
Yahoo!LDA~\cite{Ahmed12} & 11.8 hr & N/A & N/A & N/A \\ \hline 
%GraphLab~\cite{Gonzalez12} & N/A & N/A & N/A & N/A \\ \hline
\hline\end{tabular}}
{\\ \hfill (*terminated by the cluster)} \vspace{-.2cm}    
\end{table}

\begin{figure}[t]
\centering
\subfigure[]{
\includegraphics[width=0.22\textwidth]{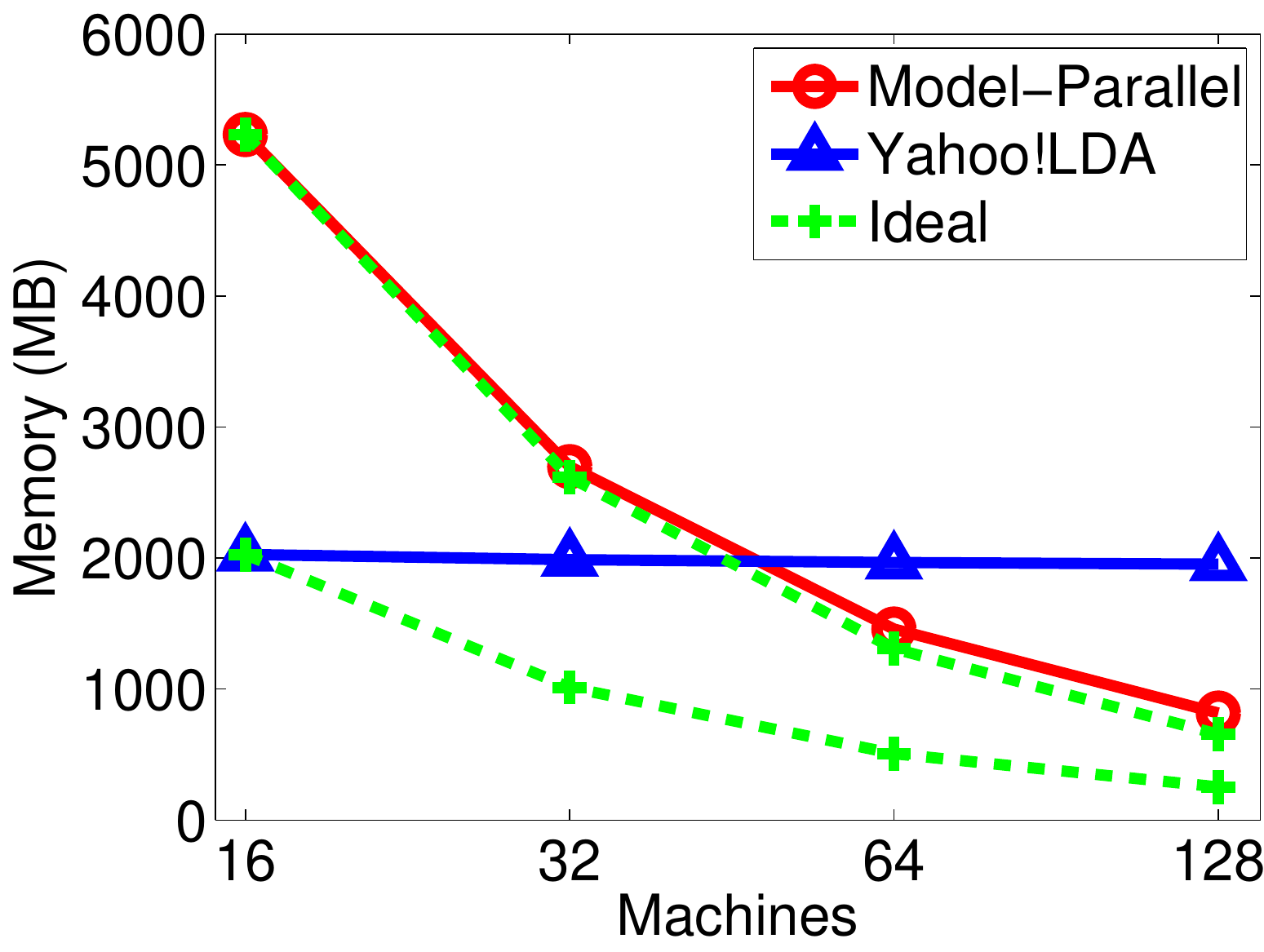}
} 
\subfigure[]{
\includegraphics[width=0.22\textwidth]{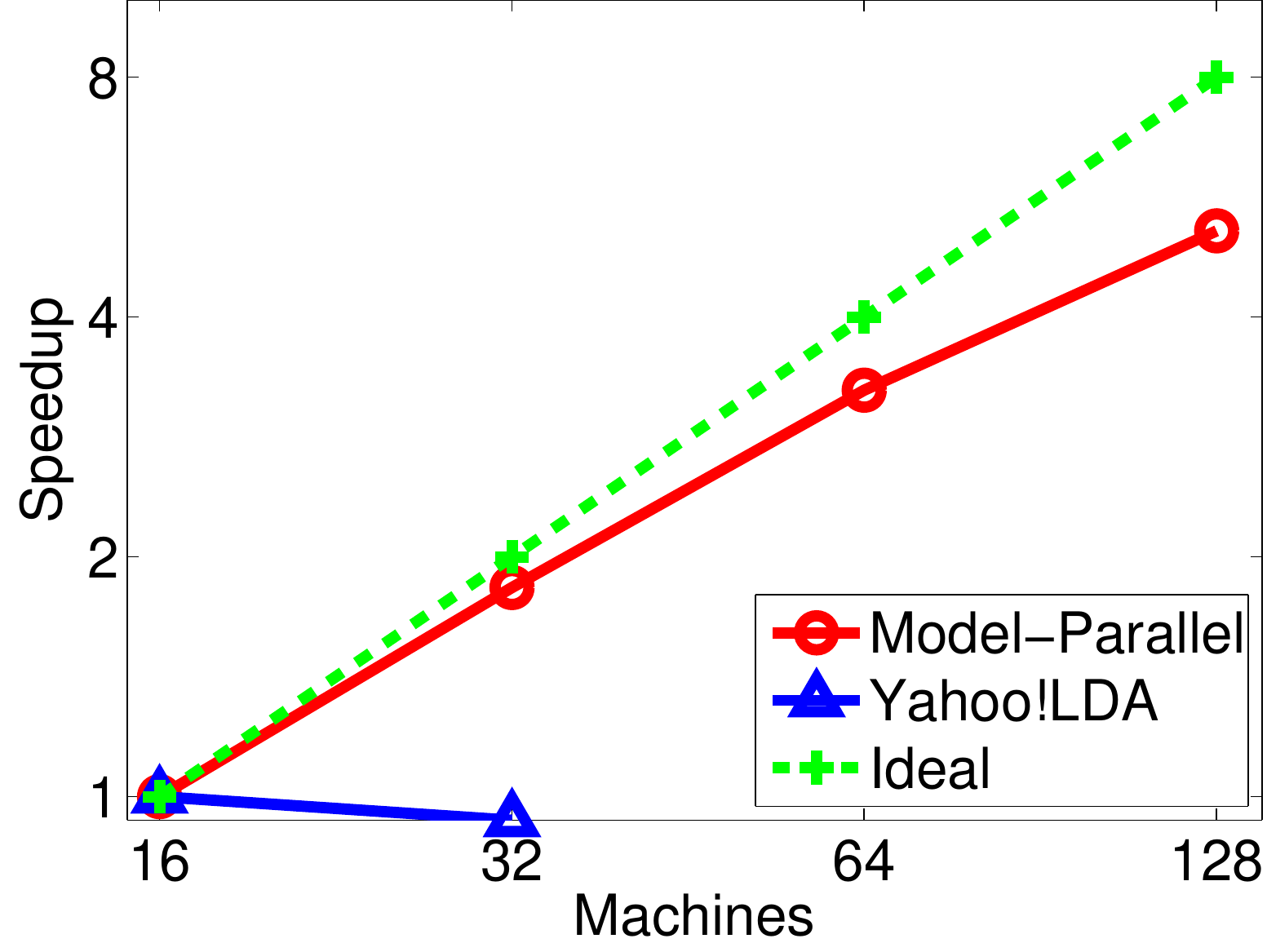}
} 
\caption{(a) Memory usage per machine as a function of the number of machines used.
Our method follows a $1/M$ trend, indicating efficient partitioning of 
both data and models across machines. 
(b) Speedup in terms of time to reach the log-likelihood of
$-2.7 \times 10^9$ on different number of machines. 
Our method achieves nice scalability, whereas Yahoo!LDA fails to utilize more computing resources.
}
\label{fig:speed}
\end{figure}

\subsection{Scalability}
In Figure \ref{fig:speed}(a), 
we show the total memory footprint of each worker as the number of computing nodes increase, using unigram dataset with $ K=5000 $ topics.
In the ideal case, as the number of machines doubles, the memory consumption should be halved. 
We can observe that the model-parallel inference achieves nearly ideal scalability over machines. 
Although starts with a higher memory footprint, it closely follows a $1/M$ trend and drops to a much lower number, indicating the dynamic model partitioning scheme
effectively makes use of more memory storage without unnecessary duplication; whereas Yahoo!LDA's per-machine memory usage is almost constant, again because its data-parallel strategy requires
most of the word-topic table $\{C_k^t\}$ to be stored on each machine, indicating that adding machines will not solve big model problems.

We also show convergence speedup as a function of number of machines. 
In Figure \ref{fig:speed}(b), we show the speedup in terms of convergence time on different number of machines
for a fixed model size (unigram dataset with 5000 topics).
Interestingly, we can observe that Yahoo!LDA performs worse given 32 machines. 
The reason can be explained by the network congestion in the low-end cluster: since the models are frequently synchronized between every node, network traffic is increased in $ O(M^2) $.
Thus parameters are more likely to be out-of-date when increasing number of nodes given low bandwidth. 
This introduces more error to the overall procedure.
By contrast, we can see the curve for model-parallel inference follows the ideal speedup trend closely. 
This shows the model-parallel inference effectively utilizes additional computational resources without significant overhead.
Unlike full $ O(M^2) $ connections, on-demand communication strategy in model-parallel inference greatly reduces the traffic by managed synchronization, while providing guarantee for model correctness.
This demonstrates the ability of model-parallel inference to handle large scale inference problems on low-end clusters.

%% file: conclusion.tex
%!TEX root = main.tex

\section{Conclusion}\label{sec:end}
In this paper, we presented a model-parallel inference for LDA, 
motivated by the pitfalls of data-parallelism in distributed inference.
We proposed a system that implements {\em model-parallelism} on top of data-parallelism, and show empirical results on improved time and memory efficiency over other approaches.
In a word, model-parallelism not only eliminates dependency between inference processes but also brings the capability of handling big models.
Therefore without drastic change in the algorithm itself, e.g., using crafted Metropolis-Hasting to speed up the sampler, we can already improve the algorithm significantly just by careful arrangement of model blocks.

We expect the idea can be applied to a broader class of models as well to scale up without sophisticated algorithmic tweak.
The first attempt to the generalized model-parallelism can be found in \cite{Lee14}, which nonetheless deserves further investigation.
We are also interested in employing model-parallelism in more challenging tasks, for example Bayesian nonparametric models like Hierarchical Dirichlet Process (HDP)~\cite{Teh06} and regularized Bayesian models such as MedLDA~\cite{Zhu12}.